\documentstyle[prl,aps,multicol,epsf]{revtex}
\begin{document}
\widetext

\title{Phase Transition in the Two-Dimensional Gauge Glass} 
\author{M.Y. Choi and Sung Yong Park}
\address{Department of Physics and Center for Theoretical Physics\\
     Seoul National University\\
     Seoul 151-742, Korea}
\maketitle
\draft

\begin{abstract}
The two-dimensional XY gauge glass, which describes disordered
superconducting grains in strong magnetic fields, is investigated, 
with regard to the possibility of a glass transition.
We compute the glass susceptibility and the correlation function 
of the system via extensive numerical
simulations and perform the finite-size scaling analysis. 
This gives strong evidence for a finite-temperature transition, 
which is expected to be of a novel type.
\end{abstract}

\thispagestyle{empty}


\begin{multicols}{2}
The gauge glass model, which was originally proposed as a generalization of 
the spin-glass model~\cite{H}, has attracted much attention 
in relation to the vortex-glass phase of high-$T_c$ 
superconductors~\cite{FFH}.
In three dimensions, the XY gauge glass model is believed to
exhibit a finite-temperature glass transition~\cite{HS,Cieplak},
in agreement with experimental evidence for the vortex-glass phase at
finite temperatures \cite{exp}. 
In two dimensions, on the other hand, there has been controversy as to the
existence of a finite-temperature transition.
Equilibrium studies of 
several quantities such as the defect-wall energy~\cite{G} 
and the root-mean-square current~\cite{RY} have suggested the absence of
ordering at finite temperatures; 
this appears to be consistent with experiment, where no evidence 
for the glassy phase at finite temperatures has been found~\cite{DWKHG}.
Indeed the gauge-glass order parameter has been shown analytically 
to be zero at any finite temperature~\cite{N2}. 
However, the helicity modulus computed via Monte Carlo simulations indicates
a signal of a glassy phase at low but finite temperatures~\cite{Jeon}.
Dynamical simulations also give conflicting results: 
Whereas earlier study of 
the current-voltage characteristics has shown a strong evidence for the 
possibility of glass ordering at finite temperatures~\cite{L}, later
one has been interpreted to be consistent with the zero-temperature 
transition~\cite{Hyman}.
However, the lowest temperature considered in Ref.~\cite{Hyman} is
apparently higher than the estimated transition temperature $T_c \approx 0.15$
(in units of the coupling energy $J$) \cite{L}.  
Indeed the recent study of relaxation dynamics has indicated a finite
transition temperature $T_c \approx 0.22$ \cite{Kim}.
It should be noted here that 
the absence of an ordered phase does not necessarily imply the absence 
of a phase transition, as we will discuss later.
We thus believe that the presence/absence of a finite-temperature 
transition in two-dimensions as well as its nature is still inconclusive.
For the resolution of this, careful analysis of the behavior 
should be performed at sufficiently low temperatures. 

In this work we investigate the two-dimensional XY gauge glass via 
extensive numerical calculations, with particular attention to the 
possibility of a finite-temperature glass transition. 
We adopt the equilibration test method in Refs.~\cite{FSS,BY} to obtain 
the equilibrium and to determine the equilibration time, 
up to the system size $L = 48$.
>From the obtained equilibrium configuration, we compute the glass 
susceptibility and examine it by the finite-size scaling analysis~\cite{FSS}.
This reveals remarkable divergent behavior at low but non-zero temperatures,
providing strong evidence for the finite-temperature transition.
It is discussed how the presence of such a finite-temperature transition 
can be reconciled with the results of existing studies. 

We consider the standard XY gauge glass model on a square $L \times L$ 
lattice, which is described by 
the Hamiltonian
\begin{equation} \label{Ham1}
  H = - J \sum_{\langle i,j \rangle} 
    \cos (\phi_i - \phi_j - A_{ij}),
\end{equation}
where $J$ is the coupling energy between nearest-neighboring grains, 
$\phi_{i}$ is the phase of the order parameter of the grain
at site $i \equiv (x,y)$\, $(x,y=1, 2, \cdots , L)$, 
and the bond angle $A_{ij}$'s are taken to be 
quenched random variables distributed uniformly on the interval $[0, 2\pi)$.
The presence of a glass transition in the system can be conveniently 
described by the
divergence of the glass susceptibility, given by $\chi_G =\sum_j G_{ij}$
with the correlation function of the glass order parameter
\begin{equation}
 G_{ij}\equiv
 \left[\left|\left\langle e^{i(\phi_i-\phi_j)} \right\rangle \right|^2\right],
\label{corr}
\end{equation}
where $[\cdots]$ denotes the disorder average and $\langle\cdots\rangle$ 
the thermal average.
In the limit the distance between the two grains $i$ and $j$ becomes
large, the correlation function reduces to $q^2$, the square of the 
Edwards-Anderson glass order parameter
\begin{equation}
 q\equiv\left[\left|\left\langle e^{i\phi_i} \right\rangle\right|^2\right].
\label{EA}
\end{equation}

In an infinite system the glass susceptibility is expected to display 
the critical behavior
\begin{equation}
\chi_{G} \sim \left (T-T_c\right)^{-\gamma}
\end{equation}
with the scaling relation among the exponents
\begin{equation}
\gamma=(2-\eta) \nu,
\label{scaling law}
\end{equation}
where $\eta$ describes the power law decay of the correlation at $T_c$
and $\nu$ the divergence of the correlation length $\xi$:
\begin{equation}
\xi \sim (T-T_c)^{-\nu},
\label{xi}
\end{equation}
as the temperature $T$ approaches $T_c$. 
In a system of size $L$, the glass susceptibility 
has the finite-size-scaling form~\cite{FSS,BY}
\begin{equation}
\chi_{G}=L^{2-\eta} \bar{\chi} \left( L^{1/\nu} \left(T-T_c\right)\right)
\label{finite}
\end{equation}
with the appropriate scaling function $\bar{\chi}$,
which can be examined via extensive simulations at various temperatures
and sizes.

In the simulation concerning equilibrium properties,
equilibration is an important issue. 
We follow Ref. \cite{BY} to have a criterion for equilibration, 
and consider two replicas $\alpha$ and $\beta$ of
the system with the same realization of disorder.
The susceptibility can be calculated from the overlap between the two 
replicas, according to
\begin{equation}
\chi_G(t_0)=\frac{1}{N t_0}
   \left[\sum^{t_0}_{t=1}
   \left|\sum_j e^{i\left(\phi_j^{\alpha}(t_0+t)-\phi_j^{\beta}(t_0+t)\right)}
   \right|^2\right], 
\label{eq:sus1}
\end{equation}
where $N\equiv L^2$ is the number of spins and time $t$ is measured in
units of the Monte Carlo sweep (MCS).
Note also that the time-dependent four-spin-correlation function
defined as
\begin{equation}
\chi_G(t_0)=\frac{1}{N}
   \left[\left|\sum_j e^{i\left(\phi_j(t_0)-\phi_j(2t_0)\right)} 
        \right|^2\right]
\label{eq:sus2}
\end{equation}
converges to the glass susceptibility $\chi_G$ in the limit $t_0\rightarrow
\infty$. 
Thus the equilibration time $\tau$ can be estimated 
as the value of $t_0$ at which the two expressions, Eqs.~(\ref{eq:sus1})
and (\ref{eq:sus2}), give coincident results.
The expected behavior of the equilibration time $\tau$ is
\begin{equation}
\tau \propto \xi^z \propto (T-T_c)^{-\nu z} \label{tau}
\end{equation}
in an infinite system near $T_c$, with the dynamical exponent $z$; 
in a finite system at $T=T_c$, it is expected
\begin{equation}
\tau \propto L^z. \label{tau1}
\end{equation}
Equations (\ref{tau}) and (\ref{tau1}) naturally lead to the finite-size
scaling form~\cite{Domb}
\begin{equation}
\tau = L^z \bar{\tau}( L^{1/\nu} (T-T_c)). \label{scaling tau}
\end{equation}

To examine the behavior of the glass susceptibility,
we have performed extensive simulations at several temperatures,
ranging from $T=0.2$ to $T=1.0$.
After the equilibration time $\tau$ estimated as above,
the glass susceptibility has been measured according to Eq.~(\ref{eq:sus1})
except for that the data have been taken
during sufficiently large time intervals, from three to ten times of 
the equilibration time.
Namely, $t_0$ has been chosen to be from $3\tau$ to $10\tau$,
depending on the size.
We have also performed independent runs with up to 16000
different disorder configurations,
over which the disorder average has been taken. 
Since the different disorder realizations give statistically independent
thermal averages, the statistical error can be estimated from the standard 
deviation of the results for different samples.
Here the number of the disorder realizations to get sufficiently reliable 
statistics for the equilibration time turns out much larger than that for 
the glass susceptibility.  Thus it has taken much more averages to obtain
reliable data for the equilibration time than those for 
the glass susceptibility.
In this way, for example, the equilibration time at $T=0.3$ has been
estimated to vary from $(7\pm1)\times 10^4$\,MCS ($L=4$) to
$(2\pm 0.1)\times 10^6$\,MCS ($L=24$);
it increases rapidly as the temperature is lowered, reaching
at $T=0.25$ the value $(1.5\pm 0.4)\times 10^6$ for $L=12$.

\begin{figure}
\epsfxsize=8cm \epsfysize=6cm \epsfbox{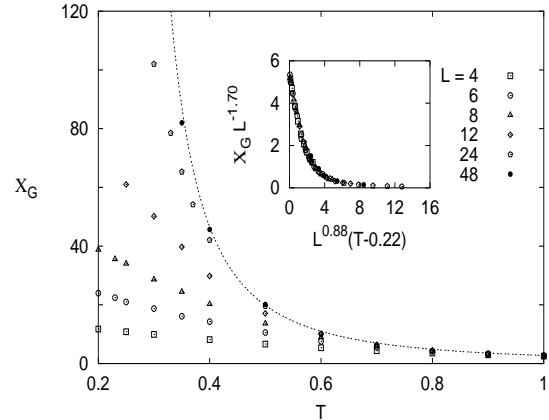}
\narrowtext
\caption{Behavior of the glass susceptibility with temperature $T$.
 The dashed line is proportional to $(T-0.22)^{-1.93}$.
 Inset shows the fit to the finite-size-scaling formula 
 given by Eq.~(\ref{finite}), with $T_c=0.22$, $1/\nu=0.88$, and
 $2-\eta=1.70$.}
\label{fig:sus}
\end{figure}

\begin{figure}
\hskip -0.4cm
\epsfxsize=8cm \epsfysize=6cm \epsfbox{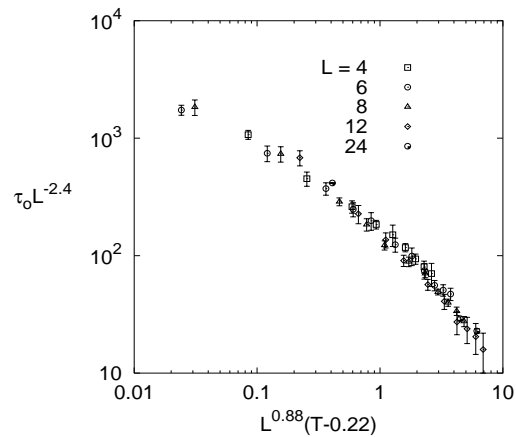}
\narrowtext
\caption{Fit of the 
obtained values of the equilibration time
to the finite-size-scaling
formula given by Eq.~(\ref{scaling tau}), with $T_c=0.22$, $1/\nu=0.88$, and
$z=2.4$.}
\label{fig:retau}
\end{figure}

Figure \ref{fig:sus} shows the behavior of the obtained glass susceptibility 
$\chi_{G}$ with the temperature $T$ for system sizes $L=4$, $6$, $8$, $12$, 
$24$, and $48$. 
In all cases, periodic boundary conditions have been employed, 
and the data points without error bars indicate that 
the errors estimated by the standard deviations are smaller than 
the size of the symbols. 
It is indeed observed that, as $L$ is increased, 
the data points apparently approach the dashed line 
representing $(T-T_c)^{-\gamma}$ with $T_c$ and $\gamma$
obtained from the finite-size-scaling method below.
The finite-size scaling analysis of the data is displayed 
in the inset of Fig.~\ref{fig:sus}, 
where the data collapse nicely to the finite-size-scaling form
given by Eq.~(\ref{finite}).
The corresponding exponents and the transition temperature are obtained:
\begin{eqnarray}
 \eta&=&0.30\pm 0.05, \nonumber\\
 \nu&=&1.14\pm 0.07, \nonumber\\
  T_c&=& 0.22 \pm 0.02,
\label{exponent}
\end{eqnarray}
which agree with the results of Ref.~\cite{Kim}.
The errors have been estimated by the size of the region 
beyond which the data do not scale well. 
The scaling law in Eq.~(\ref{scaling law}) then gives the susceptibility 
exponent
\begin{equation}
 \gamma=1.93\pm 0.08.
\end{equation}
We have also tried to fit the data to the finite-size scaling form with 
$T_c=0$,
and observed that the high-temperature data ($T\gtrsim 0.35$) appear to fit
also the $T_c=0$ scaling, in agreement with Ref.~\cite{Hyman}.
At lower temperatures, however, systematic deviation from the scaling 
has been revealed,
which becomes conspicuous for large sizes ($L=24,\,48$)~\cite{park}.
Note that such low-temperature regions, which are crucial in differentiating
the two scalings ($T_c =0$ and $T_c\neq 0$), have not been probed properly
in previous studies concluding a zero-temperature transition.
Moreover, even for small sizes such zero-temperature scaling yields $\eta$ very close to zero,
which is rather unlikely in view of the large ground-state degeneracy
expected in the system~\cite{BY}.
We also perform the finite-size scaling analysis of the equilibration time.
In this case, assuming the scaling form in Eq.~(\ref{scaling tau}),
we use the values of the exponent $\nu$ and the transition temperature $T_c$
in Eq.~(\ref{exponent}), and estimate the dynamic exponent $z$.
Figure~\ref{fig:retau} shows that such scaling is indeed reasonable,
with the dynamic exponent estimated as
\begin{equation}
 z=2.4\pm 0.3.
\end{equation}
In contrast, these equilibration time data hardly fit
the scaling with $T_c=0$, 
displaying marked deviation at $T\lesssim 0.4$~\cite{park}.

We now examine the implications of such a finite-temperature transition,
and first consider the scaling behavior of the defect-wall 
energy, which is defined to be the difference in the ground-state energy 
upon changing the boundary conditions along one direction from periodic 
to antiperiodic~\cite{Cieplak,G}.
The defect-wall energy fluctuates from sample to sample with zero mean,
and its typical value, which may be taken as the average of the absolute value
over quenched randomness, scales with the system size $L$ according to
\begin{equation} \label{defect}
  \Delta E \propto L^{\theta},
\end{equation}
in the asymptotic domain ($L \rightarrow \infty$)~\cite{Bray}.  The sign of
the exponent $\theta$ then determines the presence or absence of long-range
order at finite temperatures: Whereas for $\theta$ positive, 
the system displays rigidity or ordering, the negative value of $\theta$, 
obtained numerically in two dimensions~\cite{G},
implies ubiquity of long-wavelength fluctuations and thus
the absence of order. 
Here it should be stressed that the absence of (long-range) order does not
necessarily correspond to the absence of a finite-temperature transition,
as the Berezinskii-Kosterlitz-Thouless (BKT) transition and the associated
algebraic order present an example~\cite{KT}.
In fact it is easy to show that in any two-dimensional system described
by the Hamiltonian (\ref{Ham1}), gapless spin-wave excitations 
prevent $\theta$ from having a positive value:
Suppose that the ground-state energy in the periodic boundary conditions 
(PBC), $E_p$, 
is smaller than that in the antiperiodic boundary conditions (APBC).
>From the ground-state configuration $\{\phi_{i}^{(0)}\}$ in the PBC, 
we can make a configuration satisfying the APBC by rotating the phases 
according to $\phi_{i}^{(0)} \rightarrow \phi_{i}^{(0)} + x \pi / L$.
It is then obvious that the energy $\tilde{E}$ of the new configuration 
is not smaller than the ground-state energy $E_a$ in the same APBC.
In the limit $L\rightarrow \infty$, 
upon expanding the energy $\tilde{E}$ of the new configuration
around the ground-state configuration $\{\phi_{i}^{(0)}\}$,
we obtain the energy change (in units of $J$)
$
\tilde{E}-E_p 
=
O(L^0),
$
and thus the desired relation 
\begin{equation}
 E_a - E_p \leq \tilde{E}-E_p = O(L^0),\label{oL0}
\end{equation}
which shows that $\theta$ cannot be positive. 
For $E_a$ smaller than $E_p$, similar argument starting from the ground-state
configuration in the APBC then yields $E_p -E_a \leq O(L^0)$.
Accordingly, it may be the case that
the negative value of $\theta$ results from
spin-wave excitations, which destroy long-range order but 
maintain criticality.

To examine this possibility, we have generated 
ground states in the PBC and in the APBC
via extensive simulations for several system sizes (up to $L=24$),
which has confirmed the value of $\theta$ obtained in
the existing studies on small sizes~\cite{G}.
We have then carefully compared the two corresponding ground-state 
configurations, one in the PBC and the other in the APBC, and
found only small difference in the total number of vortices.
In particular the relative difference appears to decrease with the 
system size~\cite{park},
indicating that no (bulk) vortex excitation is involved.
This suggests that the ubiquitous fluctuations implied by the negative
value of $\theta$ are indeed of the spin-wave type rather than of the vortex 
one~\cite{Ising}.
Such spin-wave excitations should lead to the algebraic decay of
the correlation function in Eq.~(\ref{corr}) and
to the vanishing Edwards-Anderson glass order parameter 
defined in Eq.~(\ref{EA}) at all finite temperatures, thus
consistent with the result of Ref.~\cite{N2}.
It is thus strongly suggested that the system displays
quasi-long-range glass order below the finite transition 
temperature~\cite{Ozeki}, characterized by the algebraic decay of
the glass correlation function.

\begin{figure}
\vskip -1.0cm
\hskip -0.4cm
\epsfxsize=8cm \epsfysize=6cm \epsfbox{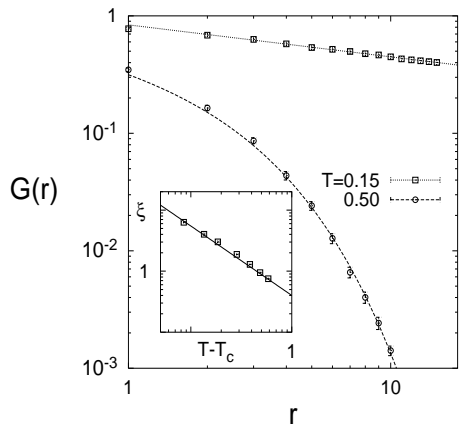}
\narrowtext
\caption{Behavior of the correlation function with distance $r$ at two 
temperatures, $T= 0.15$ and $0.50$.  Algebraic decay is obvious at $T=0.15$.
Inset shows correlation length $\xi$ versus temperature $T-T_c$ in the log-log scale
with $T_c=0.22\pm 0.01$. The slope of the soild line is fixed to 
the value $-\nu = -1.14$.
}
\label{fig:cor}
\end{figure}

We have thus computed the glass correlation function via large-scale simulations
for the system size $L=48$
and display the obtained behavior in Fig.~\ref{fig:cor},
where the difference in the behavior at the two temperatures $T=0.15$ and $0.50$ 
is spectacular:
While the dashed line at $T=0.50$ represents the least-square fit to the usual 
exponential decay $r^{-\eta} e^{-r/\xi}$ with $\eta=0.31$ and $\xi=1.90$,
the algebraic decay at $T=0.15$ is manifested by the least-square fit
(represented by the dotted line) to $r^{-\eta}$ with $\eta=0.27$.
The decrease of the exponent $\eta$ with temperature, from the value $0.30$ (at $T_c$),
reflects the criticality of the system below $T_c$,
confirming quasi-long-range glass order at low temperatures. 
The fit of the data at $T=0.15$ to the exponential-decay form has been tried, only
to yield very large deviations as well as inconsistent values of parameters $\eta$ and $\xi$.
We have also computed the correlation length at various temperatures above $T_c$ 
and found {\em perfect} scaling to the form in Eq.~(\ref{xi})
with the values of $T_c$ and $\nu$ given by Eq.~(\ref{exponent}) (see the inset).
%

We finally point out that the existence of the finite-temperature
glass transition is not inconsistent with the experimental results. 
The temperature scale in the gauge-glass model may be obtained from the
correspondence between the model temperature $T = 1.15$ (in units of $J$) 
and the real temperature $T = 12{\rm K}$ in experiment \cite{Hyman}.  
Accordingly, the transition temperature estimated here 
corresponds to approximately $2.0{\rm K}$; 
such a low-temperature regime was not probed in Ref.~\cite{DWKHG} and 
it would be interesting to investigate the low-temperature regime
experimentally. 


We would like to thank B. Kim, D. Kim, S. Ryu, 
and D. Stroud for helpful discussions, 
and acknowledge the partial support from the Korea Research Foundation
and from the Korea Science and Engineering Foundation.
The numerical works have been performed on the cray T3E at SERI
and on the SP2 supercomputer system at ERCC.



\end{multicols}


\begin{references}
\bibitem{H} J.A. Hertz, Phys. Rev. B {\bf 18}, 4875 (1978); 
   H. Nishimori, Prog. Theor. Phys. {\bf 66}, 1169 (1981).
\bibitem{FFH} D.S. Fisher, M.P.A. Fisher, and D.A. Huse, Phys. Rev. B 
   {\bf 43}, 130 (1991).
\bibitem{HS} D.A. Huse and H.S. Seung, Phys. Rev. B {\bf 42}, 1059 (1990). 
\bibitem{Cieplak}
   M. Cieplak, J.R. Banavar, and A. Khurana, J. Phys. A {\bf 24}, L145 (1991);
   J.D. Reger, T.A. Tokuyasu, A.P. Young, and M.P.A. Fisher, Phys. Rev. B
   {\bf 44}, 7147 (1991);
   M. Cieplak, J.R. Banavar, M.S. Li, and A. Khurana, {\it ibid}. {\bf 45}, 
   786 (1992).
\bibitem{exp} R.H. Koch, V. Foglietti, W.J. Gallagher, G. Koren, A. Gupta, 
   and M.P.A. Fisher, Phys. Rev. Lett. {\bf 63}, 1511 (1989); 
   P.L. Gammel, L.F. Schneemeyer, and D.J. Bishop, {\it ibid.} {\bf 66}, 953
   (1991); C. Dekker, W. Eideloth, and R.H. Koch, {\it ibid.} {\bf 68}, 3347
   (1992). 
\bibitem{G} M.J.P. Gingras, Phys. Rev. B {\bf 45}, 7547 (1992); 
   M.P.A. Fisher and T.A. Tokuyasu, Phys. Rev. Lett. {\bf 66}, 2931 (1991);
 J.M. Kosterlitz and M.V. Simkin, {\it ibid.} {\bf 79}, 1098 (1997).
\bibitem{RY} J.D. Reger and A.P. Young, J. Phys. A {\bf 26}, L1067 (1993).
\bibitem{DWKHG} C. Dekker, P.J.M. W\"oltgens, R.H. Koch, B.W. Hussey, 
   and A. Gupta, Phys. Rev. Lett. {\bf 69}, 2717 (1992).
\bibitem{N2} H. Nishimori, Physica A {\bf 205}, 1 (1994).
\bibitem{Jeon} G.S. Jeon, S. Kim, and M.Y. Choi, Phys. Rev. B {\bf 51}, 16211
    (1995).
\bibitem{L} Y.H. Li, Phys. Rev. Lett. {\bf 69}, 1819 (1992).
\bibitem{Hyman} R.A. Hyman, M. Wallin, M.P.A. Fisher, S.M. Girvin, and A.P.
    Young, Phys. Rev. B {\bf 51}, 15304 (1995).
\bibitem{Kim} B.J. Kim, M.Y. Choi, S. Ryu, and D. Stroud, 
   Phys. Rev. B {\bf 56}, 6007 (1997).
\bibitem{FSS} V. Privman, {\it Finite-Size Scaling Analysis and Numerical 
 Simulation of Statistical Systems} (World Scientific, Singapore, 1992). 
\bibitem{BY} R.N. Bhatt and A.P. Young, Phys. Rev. B {\bf 37}, 5606 (1988).
\bibitem{Domb} See, e.g., M.N. Barber, in 
{\it Phase Transitions and Critical Phenomena,} 
edited by C. Domb and J. Lebowitz, (Academic, New York, 1983), Vol. 8,
p. 146.
\bibitem{park} S.Y. Park and M.Y. Choi (unpublished).
\bibitem{Bray} A.J. Bray and M.A. Moore, J. Phys. C {\bf 17}, L463 (1984).
\bibitem{KT} J.M. Kosterlitz and D.J. Thouless, J. Phys. C {\bf 6}, 
 1183 (1973); J.M. Kosterlitz, {\it ibid}. {\bf 7}, 1046 (1974).
\bibitem{Ising} This is to be contrasted with the Ising system, 
where the ubiquitous fluctuations (for $\theta<0$) are domain walls 
and no criticality is allowed.
\bibitem{Ozeki} Y. Ozeki and H. Nishimori, J. Phys. A {\bf 26}, 3399 (1993).
\end{references}
\end{document}